\documentstyle[prc,aps,epsfig,preprint]{revtex}
\def\bm #1{ \bbox{#1} }
\def\beq{\begin{equation}}
\def\eeq{\end{equation}}
\def\veck{{\bf k}}
\def\vecr{{\bf r}}
\def\vecR{{\bf R}}

\newcommand{\be}{\begin{eqnarray}}
\newcommand{\ee}{\end{eqnarray}}

\begin{document}

\draft

\title{ Deuteron distribution in nuclear matter}
\author{O.~Benhar$^1$, A.~Fabrocini$^2$,S.~Fantoni$^{3,4}$, 
        A.Yu.~Illarionov$^5$, G.I.~Lykasov$^5$}
\address{$^1$INFN, Sezione di Roma, I-00185, Roma, Italy}
\address{$^2$Dip. di Fisica, Universit\`a di Pisa, and INFN, Sezione di 
Pisa, I-56100 Pisa, Italy}
\address{$^3$International School for Advanced Studies, SISSA , 
I-34014,Trieste, Italy}
\address{$^4$ International Centre for Theoretical Physics, ICTP, I-34014
Trieste, Italy}
\address{$^5$Joint Institute for Nuclear Research,
              141980 Dubna, Moscow Region, Russia}

\date{\today}

\maketitle

\begin{abstract}
We analyze the properties of deuteron-like structures in 
infinite, correlated nuclear matter, described by a realistic 
hamiltonian containing the Urbana $v_{14}$ two--nucleon and the 
Urbana TNI many--body potentials. The distribution 
of neutron-proton pairs, carrying the deuteron quantum
numbers, is obtained as a function of the total momentum by computing 
the overlap between the nuclear matter in its ground state 
and the deuteron wave functions in correlated basis functions theory. 
We study the differences between 
the ${\rm S}$- and ${\rm D}$-wave components of the deuteron 
and those of the deuteron-like pair in the nuclear medium. 
The total number of deuteron type pairs is computed and compared
with the predictions of Levinger's quasideuteron model. The resulting 
Levinger's factor in nuclear matter  at equilibrium density 
is 11.63.  We use the local density approximation 
to estimate the Levinger's factor for heavy nuclei, obtaining results
which are consistent with 
the available experimental data from photoreactions.  
\end{abstract}

\pacs{PACS number(s): }

\section{Introduction}

The suggestion that the nuclear response may be interpreted as the response of 
a collection of neutron-proton ($np$) pairs carrying the quantum numbers of the 
deuteron was first put forward in the fifties by Levinger \cite{Lev51} 
and Gottfried \cite{Got58}, to explain nuclear photoabsorption data.

The basic idea underlying the Levinger's {\it quasideuteron} (QD) model is 
that the nuclear photoabsorption 
cross section $\sigma_A(E_\gamma)$, 
above the giant dipole resonance and below the pion threshold, 
is proportional to that corresponding to the break--up 
of a deuteron embedded in hadronic matter, and denoted hereafter as  
$\sigma_{QD}(E_\gamma)$ 

\beq 
\sigma_A(E_\gamma) = {\cal P}_D \  
\sigma_{QD}(E_\gamma)\ .
\label{levinger:formula}
\eeq
The proportionality constant ${\cal P}_D$ has to be interpreted
as the fraction of the $A(A-1)/2$ nucleon--nucleon pairs, which are of 
QD type, and it is given by

\beq
{\cal P}_D = {\rm L}\ \left[ \frac{Z({\rm A-Z})}{{\rm A}} \right]\ , 
\label{levinger1}
\eeq
where $A$ and $Z$ denote the nuclear mass and charge
and $L$ is the so called Levinger's factor. 
${\cal P}_D$ can be directly calculated from the ground state wave function
of the nucleus with mass $A$. Since the deuteron is a bound state, 
${\cal P}_D$ scales with the number of particles $A$.

 From ${\cal P}_D$, the probability of finding a 
deuteron-like nucleon pair in a complex nucleus can be easily extracted. 
Such probability can be obtained by normalizing the number
of QD pairs ${\cal P}_D$ to the total number of pairs, and, 
therefore, it is inversely 
proportional to the number of particles. The probability is zero in 
infinite nuclear matter, unless the nuclear matter wave function 
contains a long range order, providing a condensation of QD pairs. 
 
According to the Levinger's model\cite{Lev79} $\sigma_{QD}(E_\gamma)$ is 
taken as the deuteron
cross section times a damping function, of exponential form, 
accounting for the Pauli blocking of the final states available 
to the nucleon ejected from the QD:  

\beq 
\sigma_{QD}(E_\gamma) = \sigma_d(E_\gamma)\ {\rm e}^{-(D/{E_\gamma})}\ .
\label{levinger2}
\eeq

Subsequently, Laget\cite{Laget81} proposed to associate  
$\sigma_{QD}(E_\gamma)$ with the transition
amplitudes of virtual ($\pi\ +\ \rho$)--meson exchanges between the two
nucleon of the QD pair,  
leading to a cross section denoted $\sigma_d^{exch}(E_\gamma)$. 

Both models fit reasonably well the existing photoreaction data in heavy 
nuclei, but the resulting factors, $L_{Lev}(A)$ and $L_{Laget}(A)$, have
different phenomenological values, with $L_{Laget}(A)$ being about
$20\%$ larger than $L_{Lev}(A)$. 
  
A generalization of the QD model was proposed by 
Frankfurt and Strikman \cite{Frank81}, to explain the production of
 fast backward protons in semi-inclusive processes off nuclear targets.
According to the model of Refs.\cite{Frank81}, generally referred 
to as {\it few nucleon correlation} model, the structure of the 
nuclear wave function at short internucleon distances is dominated 
by strongly correlated multinucleon clusters. A quantitative understanding
of the above reaction processes requires a microscopic calculation of the 
quasideuteron distribution $P_D(\veck_D)$ 
in the nucleus, as a function of its momentum $\veck_D$. Moreover,
the integral of $P_D(\veck_D)$ over $\veck_D$, being proportional to 
${\cal P}_D$, provides an unbiased calculation of the Levinger's factor
$L$. 

More recently, the occurrence and spacial structure of deuteron-like 
configurations in light nuclei has been studied using the Green's Function 
Monte Carlo (GFMC) method \cite{For96}. It is interesting to extend such 
analysis to heavier nuclei and to nuclear matter.

Systematic quantitative investigations of nucleon-nucleon (NN) correlations in 
nuclear matter
have been carried out within microscopic many-body theories 
(for a recent review see Ref.\cite{Marcelbook}). In particular, Correlated 
Basis Function (CBF) theory has been 
applied to obtain the nuclear matter momentum distribution \cite{Fan84} and 
spectral functions \cite{Ben89,Ben92,Ben94,Ben00} from realistic hamiltonians.
In this paper we use the same many body framework
to carry out an {\it ab initio} calculation of the momentum distribution
$P_D(\veck_D)$ 
of QD pairs in infinite nuclear matter, as well as of the associated total 
number of QD pairs per particle ${\cal P}_D/A$.
 
The definition of the QD total momentum distribution in terms of the overlap 
between the nuclear matter and the deuteron ground state wave functions is 
given in Sec. II, where the many-body formalism employed in the calculations 
is also briefly outlined. In Sec. III the results of numerical 
calculations, including both the QD momentum distribution and ${\cal P}_D$
in nuclear matter at the empirical saturation density, $\rho=$ 0.16 fm$^{-3}$, 
are discussed and compared to the empirical estimates of the Levinger's
factor. Finally, the summary and conclusions are given in Sec. IV.

\section{Formalism}

The distribution of QD pairs with total momentum $\veck_D$
in nuclear matter is defined as
(sum over repeated greek indeces is implicit hereafter)
\beq
P_D(\veck_D) = \frac{1}{2J_D+1} \sum\limits_{i<j} \sum_{n} \left| 
M_{ij}^{n \alpha}(\veck_D) \right|^2\ ,
\label{def:P_D}
\eeq    
where $J_D=1$ is the spin of the deuteron, and  
\beq
M_{12}^{n \alpha}(\veck_D) = \int d{\widetilde R} d^3r_1 d^3r_2
\Psi_{NM}^\ast({\bf r}_1,{\bf r}_2,{\widetilde R}) 
 \Psi_D^\alpha({\bf r}_1,{\bf r}_2) \Phi_n({\widetilde R})\ ,
\label{def:P_D2}
\eeq 
with ${\widetilde R} \equiv ({\bf r}_3,\ldots,{\bf r}_A)$. In the above 
equation, $\Psi_{NM}$ and $\Phi_n$ denote the normalized nuclear matter 
ground state wave 
function and the wave function of the $({\rm A}-2)$-nucleon system in the
state $n$, respectively. The configuration space deuteron wave function (DWF) can be cast in the 
form
\beq
\Psi_D^\alpha(\vecr_{ij},\vecR_{ij}) = \frac{
{\rm e}^{i \veck_D \cdot \vecR_{ij}}}{\sqrt{\Omega}}
 \psi^\alpha_D(ij)|00\rangle \ ,
\eeq
where $\Omega$ is the normalization volume, 
$\vecR_{ij}=(\vecr_i+\vecr_j)/2$, $\vecr_{ij}=\vecr_i-\vecr_j$, 
$|00\rangle$ is the spin-isospin singlet two-nucleon state and 
the relative motion of the pair is described by
\beq
\psi^\alpha_D(ij) = \left[ u_D(r_{ij})\sigma_i^\alpha -
 {w_D(r_{ij})\over\sqrt2}T^{\alpha\beta}(\widehat{\vecr}_{ij})\sigma_i^\beta
 \right]\  .
\label{DWF:coord}  
\eeq      
 In Eq.(\ref{DWF:coord}), 
$u_D(r)$ and $w_D(r)$ are the $\ell=0$ and $\ell = 2$ components of
the deuteron wave function, normalized according to 
\beq
\int_0^\infty r^2 dr \left[ u^2(r) + w^2(r) \right] = 1\ ,
\label{uw:norm}
\eeq
$\sigma_i^\alpha$ ($\alpha = 1,2,3$) 
denote the Pauli matrices and the tensor operator is given by
\beq
T^{\alpha\beta}(\widehat{\vecr}_{ij}) =
 3\widehat{\vecr}_{ij}^\alpha\widehat{\vecr}_{ij}^\beta 
- \delta^{\alpha\beta}\ .
\label{tens:op}
\eeq

In CBF theory $|\Psi_{NM}\rangle$ is usually written, in coordinate space, 
in the form (R$\equiv({\bf r}_1,\ldots,{\bf r}_A)$ specifies the nucleon 
positions)
\beq
\Psi_{NM}(R) = {\cal S} \Bigl[ \prod_{i<j} F(ij) \Bigl] \Phi_0(R)\ ,
\label{def:NMWF}
\eeq
where ${\cal S}$ is the symmetrization operator and $\Phi_0$ is
the Slater determinant describing a noninteracting Fermi gas of nucleons
carrying momenta $\veck$ with $|\veck|\le k_F = (6 \pi^2 \rho/\nu)^{1/3}$, 
$\nu$ being the degeneracy of the momentum states (in symmetric nuclear matter 
$\nu = 4$). The operator $F(ij)$, accounting for the correlation structure 
induced by 
the nucleon nucleon (NN) interaction, has been chosen of the form
\cite{Pan79}
\be
F(ij) & = & f_c(r_{ij}) + f_\sigma(r_{ij}) (\bm\sigma_i\cdot\bm\sigma_j)
 + f_\tau(r_{ij}) (\bm\tau_i\cdot\bm\tau_j)
 + f_{\sigma\tau}(r_{ij}) (\bm\sigma_i\cdot\bm\sigma_j)( \bm\tau_i\cdot\bm\tau_j)
 \nonumber \\
 & + & f_t(r_{ij})T_{\alpha\beta}(\widehat{\vecr}_{ij})
\sigma_i^\alpha \sigma_j^\beta
+ f_{t\tau}(r_{ij})T_{\alpha\beta}(\widehat{\vecr}_{ij})
\sigma_i^\alpha\sigma_j^\beta
               (\bm\tau_i\cdot\bm\tau_j)\ ,
\label{corr.op.-coord.}
\ee
where $f_c(r), f_\sigma(r), f_\tau(r), f_{\sigma\tau}(r), f_t(r)$ and 
$f_{t\tau}(r)$ are correlation functions whose radial shapes 
are determined minimizing the expectation value of the hamiltonian 
in the ground state described by Eq.(\ref{def:NMWF}) \cite{Pan79}. 
As $r \rightarrow \infty$, $f_c(r) \rightarrow 1$, while all other 
correlation functions go to zero.

Summation over $\veck_D$ of $P_D(\veck_D)$ yields ${\cal P}_D/(2J_D+1)$
in nuclear matter. This  
number, which corresponds to an  extensive quantity and therefore is
propotional to $A$, leads to a direct evaluation of the 
Levinger's factor $L$, to be compared with the value resulting from the 
phenomenological analyses\cite{Ang86,Carlos82} 
of the available experimental data on photoreactions\cite{Lepretre81,Ahrens75}.

The quantity defined by Eqs.(\ref{def:P_D}) and (\ref{def:P_D2}) 
is related to the fully linked part of the two--nucleon density matrix,  
$\rho^{(2)}(\vecr_{1},\vecr_{2},\vecr_{1^\prime},\vecr_{2^\prime})$. 
This part is the only one providing extra information on the N--N 
correlations with respect to that carried by the one--body density matrix,
 or, equivalently, by the nucleon momentum distribution \cite{Ben00}. 
Using standard cluster expansion techniques \cite{Fan98}, 
$P_D(\veck_D)$ can be written as a series of terms involving 
an increasing number of particles. 
We have calculated the cluster contributions 
associated with the diagrammatic structure shown in fig.\ref{f1}, and 
its exchange counterpart, where the deuteron wave function $\Psi_D(1,2)$ is
multiplied by the correlation operator $F(1,2)$. This corresponds to a 
{\it dressed} leading order approximation, whose validity has been checked in 
previous CBF calculations of the response function and 
of the spectral function of nuclear matter, and whose expression reads
\beq
P_D(\veck_D) = \frac{1}{2}\frac{\rho^2}{4\pi} \int d^3r_{11^\prime} d^3r_{12} 
d^3r_{1^\prime 2^\prime}\ \mbox{e}^{i \veck_D \cdot (\vecr_{11^\prime} + 
\vecr_{22^\prime})/2}
 n\left(r_{11^\prime}\right) n\left(r_{22^\prime}\right)
 \ \Sigma(\vecr_{12},\vecr_{1^\prime 2^\prime})\ ,
 \label{pD:coord}
\eeq
with
\beq
\Sigma(\vecr_{12},\vecr_{1^\prime 2^\prime}) = \frac{1}{3}
 Tr\left[ F^\dagger (1^\prime 2^\prime) 
 \psi_D^{\alpha \dagger}(1^\prime 2^\prime)
\Pi_{00} \psi_D^\alpha(12) F(12)
 \left(1 - P_\sigma P_\tau\right) \right]\ .
 \label{def:sig}
\eeq
In the above equation, $\Pi_{00}$ is the operator projecting onto the 
$S=0$, $T=0$ two-nucleon state:
\beq
\Pi_{00} = \frac{1 - (\bm\sigma_1\cdot\bm\sigma_2)}{4} \
          \frac{1 - (\bm\tau_1\cdot\bm\tau_2)}{4}\ ,
\eeq
while the spin- and isospin-exchange operators, $P_\sigma$ and  $P_\tau$, 
are given by
\beq
P_\sigma = \frac{1 + (\bm\sigma_1\cdot\bm\sigma_2)}{2},\ \ \ 
 P_\tau = \frac{1 + (\bm\tau_1\cdot\bm\tau_2)}{2}\ .
 \label{exch.op.}
\eeq

The function $n(r)$ is the {\it correlated} one-body density 
matrix \cite{Fan84}, normalized as $n(r=0)=1$,  
and trivially related to the nucleon momentum distribution,
 $n(k)$, through
\beq
n(r) = \frac{\nu}{(2\pi)^3\rho}\int\ d^3k \mbox{e}^{i\veck \cdot \vecr} 
n(k)\ .
\label{nk:norm}
\eeq

Evaluation of the trace appearing in Eq.(\ref{def:sig}) leads to 
the simple result 
\beq
\Sigma(\vecr,\vecr^\prime) =  \frac{1}{16}\left[ U(r)U(r^\prime) +
 W(r)W(r^\prime) Q(\widehat{\vecr}, \widehat{\vecr}^\prime)\right] \ ,
\label{sum} 
\eeq 
where
\beq    
Q(\widehat{\vecr}, \widehat{\vecr}^\prime) = \frac{1}{2}
\left[ 3\left(\widehat{\vecr}\cdot \widehat{\vecr}^\prime\right)^2 - 
1\right]\ ,
\label{def:Q}
\eeq
and the functions $U(r)$ and $W(r)$ are defined as
\beq
U(r) = u_D(r) - \Delta u(r)\ , 
\label{def:U-coord} 
\eeq
\beq
W(r) = w_D(r) - \Delta w(r)\ .
\label{def:W-coord}      
\eeq
The explicit expression of the functions
$\Delta u(r)$ and $\Delta w(r)$, yielding the deviation of 
$U(r)$ and $W(r)$ from the bare components of the DWF, are 
\be
\nonumber
\Delta u(r) &=& 
 u_D(r)\left[ h_c(r) - f_\sigma(r) + 3f_\tau(r) +
3f_{\sigma\tau}(r) \right] \\
 & - & \sqrt{8}w_D(r)\left[f_t(r) - 3f_{t\tau}(r) \right]\ ,
\label{def:DU-coord} 
\ee
and
\be
\nonumber
\Delta w(r) &=& w_D(r)\left[ h_c(r) - f_\sigma(r) + 3f_\tau(r) +
3f_{\sigma\tau}(r) \right] \\
 & - & \sqrt{8}\left(u_D(r) - \frac{w_D(r)}{\sqrt{2}} \right)
   \left[f_t(r) - 3f_{t\tau}(r) \right]\ ,
\label{def:DW-coord}
\ee
with $h_c(r) = 1 - f_c(r)$. Note that in absence of correlations, 
i.e. setting $f_c(r) \equiv$1 and all other correlation functions 
identically equal to zero, $U(r)$ and $W(r)$ reduce to $u_D(r)$ and
$w_D(r)$, respectively.

Using the functions defined in Eqs.(\ref{def:U-coord}) and 
(\ref{def:W-coord}), the wave function 
describing the motion of the QD pair in nuclear matter can be 
written in the same form as the DWF (see Eqs.(\ref{def:P_D}), (\ref{def:P_D2})
and (\ref{DWF:coord})):
\beq
\Psi_{QD}^\alpha(\vecr_{ij},\vecR_{ij}) = 
\frac{ {\rm e}^{i \veck_D \cdot \vecR_{ij}} }{\sqrt{\Omega}} 
\left[ U(r_{ij})\sigma_i^\alpha - \frac{W(r_{ij})}{\sqrt{2}}
T^{\alpha\beta}(\widehat{\vecr})\sigma_i^\beta  
\right] |00\rangle \ .  
\label{def:dwfnm}
\eeq

Using the Fourier tranforms of $U(r)$ and $W(r)$, defined as
\beq
U(k) = \sqrt{ \frac{2}{\pi} }
       \int_0^\infty r^2dr j_0(kr) U(r)\ ,
\label{def:U-mom} 
\eeq
and
\beq
W(k) = \sqrt{ \frac{2}{\pi} }
       \int_0^\infty r^2dr j_2(kr) W(r)\ ,
\label{def:W-mom}
\eeq
$j_0(kr)$ and $j_2(kr)$ being spherical Bessel functions, and the nucleon 
momentum distribution in nuclear matter, $n(k)$, Eq.(\ref{pD:coord}) can 
be rewritten in the form
\beq
P_D(\veck_D) = \int d^3k\ P(\veck_D,\veck)\ ,
\label{pD:basic-mom}
\eeq
where
\beq
P(\veck_D,\veck) =  
 n\left(\left| \frac{\veck_D}{2} - \veck \right| \right)\ 
 n\left(\left| \frac{\veck_D}{2} + \veck \right| \right)
|\Psi_{QD}(k)|^2\ ,
\label{pD1}
\eeq
and
\beq
|\Psi_{QD}(k)|^2\ = \frac{1}{4\pi}\left[U^2(k)+W^2(k)\right]\ .
\label{PSID}
\eeq

The above equations have been used to carry out the numerical calculations.

It has to be noticed that the contributions arising from the 
non commuting structure of the correlations reaching the four
external vertices, $1$, $2$, $1'$ and $2'$, of the diagrammatical 
structure of fig. \ref{f1}, are not exactly accounted for, 
but only according to the {\sl dressed} leading order approximation.

\section{Results}

Fig. \ref{f2} shows the behavior of $U(r)$ and $W(r)$ evaluated using 
a many body hamiltonian including the Urbana $v_{14}$ NN potential and
supplemented by the TNI model of many-body forces \cite{Lag81}. 
For comparison, we also show the components of the Urbana $v_{14}$ DWF and the 
functions $\Delta u$ and $\Delta w$ defined in Eqs.(\ref{def:DU-coord}) 
and (\ref{def:DW-coord}), respectively.
It appears that the main differences
between deuteron and QD occur at $r <$ 2 fm.
At small relative distance ($r <$ 1 fm), the effect of the nuclear medium 
leads to an appreciable suppression of $U(r)$ with respect to $u_D(r)$, 
whereas $W_D(r)$ turns out to be substantially enhanced, 
compared to $w_D(r)$. 

The momentum space behavior of $|U(k)|$, $|W(k)|$, $|u_D(k)|$, 
$|w_D(k)|$, $|\Delta u(k)|$ and $|\Delta w(k)|$ is displayed in fig. \ref{f3}. The 
main effect of the nuclear medium appears to be a shift of the second 
mimimum of both $|U(k)|$ and $|W(k)|$ towards lower values of k.

Eqs.(\ref{def:DU-coord}) and (\ref{def:DW-coord})
show that the nuclear medium modifications to the DWF are driven by 
the functions $H_t(r) = f_t(r) - 3f_{t\tau}(r)$ and 
$\Delta H_c(r) = -f_\sigma(r) + 3f_\tau(r) + 3f_{\sigma\tau}(r)$, 
resulting from the combination of different components of the NN
correlation operator. The radial dependence of $H_t(r)$ and 
$\Delta H_c(r)$, illustrated in fig. \ref{f4}, shows that the effect of
scalar and spin-isospin correlations, described by $\Delta H_c(r)$, dominates 
at very short relative distance, whereas $H_t(r)$, accounting for tensor
correlations, has a significantly longer range. 

The distribution of deuteron pairs with total momentum $\veck_D$, 
$P_D(\veck_D)$, resulting
from our approach is displayed in fig. \ref{f5} as a solid line. 
Within the Fermi gas model, $P_D(\veck_D) \equiv$ 0 at 
$|\veck_D| > 2 k_F$, 
implying that the high momentum tail of $P_D(\veck_D)$ is entirely due to NN 
correlations. The distribution of deuterons in a Fermi gas is represented 
by a dashed line in the figure. 
The comparison between the two curves 
clearly shows that the correlations deplete the distribution with 
respect to the Fermi gas at $|\veck_D| < 2 k_F$. The depletion 
is mostly due to the non central, tensor correlations.  

Similarly, one can define 
the relative momentum distribution of the nucleons belonging to a QD pair 
in nuclear matter 

\beq
P^{\it rel}_D(k) = \phi(k)|\Psi_{QD}(k)|^2\ , 
\label{def:Phid}
\eeq

where
 
\beq
\phi(k) = \int d^3k_D\ 
 n\left(\left| \frac{\veck_D}{2} - \veck \right| \right)            
 n\left(\left| \frac{\veck_D}{2} + \veck \right| \right)\ .    
\label{def:phi}
\eeq

For example, 
in the Fermi gas model $n(k) = \theta(k_F-k)$, and $\phi(k)$ takes the
simple form 
\beq
\phi(k) = (2\pi)^3 2 \rho\ \left( 1 - \frac{3}{2}x + \frac{1}{2}x^3 \right)
\theta(1 - x)\ ,
\eeq
with $x=k/k_F$.

Fig. \ref{f6} shows the relative momentum distribution of a QD pair in 
nuclear matter, as well as the functions $|\Psi_{QD}(k)|^2$ and $\phi(k)$ 
defined by Eqs.(\ref{PSID}) and (\ref{def:phi}), respectively. For comparison
the relative momentum distribution of a deuteron in free space is also 
displayed.

The total number of pairs of the QD type 
in nuclear matter, ${\cal P}_D$, can be 
obtained by momentum integration of either $P^{\it rel}_D(\veck)$ 
or $P_D(\veck_D)$ times the spin multiplicity, $2J_D+1=3$, of the
deuteron:
\beq
\frac{{\cal P}_D}{A} = \frac{3}{\rho}\ \int\ 
\frac{d^3k_D}{(2\pi)^3}\ P_D(\veck_D) = \frac{3}{\rho}\ 
\int\ \frac{d^3k}{(2\pi)^3}\ P^{\it rel}_D(\veck)\ .
\label{def:totprob}
\eeq
The calculation carried out using the correlated model of nuclear matter 
and Eq.(\ref{pD:basic-mom}) yields ${\cal P}_D/A = 2.895$, 
to be compared with the Fermi gas model result of $3.406$.  

In order to compare the calculated ${\cal P}_D$ to the number of QD pairs 
extracted 
from the analysis of photonuclear data we have to make a connection with 
the Levinger's formula given in Eqs.(\ref{levinger:formula}) and  
(\ref{levinger1}). The relation is given by
\beq
{\cal P}_D = {\rm L}\ \left( \frac{{\rm Z}{\rm (A-Z)}}{\rm A} \right)\ ,
\label{PD:L}
\eeq 
and, for symmetrical matter ($Z=A/2$), one has
\beq
L(A) = 4\ \frac{{\cal P}_D}{A} \ .
\eeq

The nuclear matter value resulting from our calculation gives
$L(\infty) = 11.63 $. This value should be compared with that given by 
the phenomenological formula

\beq
L_{Lev}(A) = 13.82 \frac{A}{R^3[fm^3]}\ ,
\eeq
reported in Ref.\cite{Ang86}, providing $L_{Lev}(\infty) = 9.26$. 
Notice that, for a deuteron in a Fermi gas, $L_{FG}(\infty) = 13.6$. 
Surface contributions to $L(A)$ can be estimated by 
exploiting the calculation of the enhancement factor ${\cal K}$ 
in the electric dipole sum rule for finite nuclei  
of Ref.\cite{Fabrocini85}, performed within the  
CBF theory and Local Density Approximation (LDA).  
The enhancement factor is related to experimental data
on photoreactions through the equation: 
\beq
1+{\cal K}_{exp} = 
\frac{1}{\sigma_0}\int^{m_\pi c^2} \sigma_A(E_{\gamma}) dE_{\gamma} \ ,
\label{enhanc}
\eeq
where $\sigma_0 = 60\ [Z(A-Z)/A]$ $MeV\ mb$ and $m_\pi c^2$ is the 
$\pi$--meson production treshold. 

Therefore, the Levinger's factor can be related to ${\cal K}$ in 
the mass number range where the coefficient $D$ 
in Eq.(\ref{levinger2}) is fairly $A$--independent, 
namely for sufficiently large values of $A$.
By adding the surface contributions, as extracted from Ref.\cite{Fabrocini85}, 
to  the nuclear matter bulk result, we get: 

\beq
L(A) = 11.63 - 9.76\ A^{-1/3}\ .
\label{def:mass}
\eeq

Fig. \ref{f7}  shows our results for $L(A)$ compared with 
$L_{Lev}(A)$ and $L_{Laget}(A)$, as extracted\cite{Ang86,Carlos82}   
from the available
experimental data on photoreactions. The computed Levinger's factors are 
almost A--independent for heavy nuclei ($A>100$), 
and result to be $\sim 25\%$ larger than $L_{Lev}(A)$ and 
$\sim 15\%$ smaller than $L_{Laget}(A)$, and therefore they 
are consistent with the
experimental data.  This differs from what happens for the enhancement
factor ${\cal K}$, where essentially the same theory as the one 
used in this paper leads to a value which is $\sim 60\%$ larger
than the experimental one. Therefore,
this disagreement between theory and experiment has to be mainly traced back 
to the sizeable tail contributions to the electric dipole sum rule, 
absent in the definition of Eq.(\ref{enhanc}).

\section{Conclusions}

Correlated Basis Function theory of the two--body density matrix
has been applied to compute the distribution $P_D(\veck_D)$ 
of neutron--proton pairs characterized by the deuteron wave 
function and having total momentum ${\bf k}_D$. 

It has been found that this distribution in nuclear matter is mostly
concentrated at $0\leq |\veck_D| \leq 2k_F$.
Besides being responsible for 
the appearance of the tail of $P_D(\veck_D)$ at $|\veck_D|> 2k_F$, 
NN correlations produce an appreciable effect at low momenta. The 
inclusion of correlations associated with the tensor component of the 
one pion exchange interaction leads to a $\sim$ 15$\%$ decrease of 
$P_D(\veck_D)$ at $|\veck_D|<k_F$. In general, 
inclusion of correlations reduces the prediction of the Fermi gas model 
in this region .

Summation of $P_D(\veck_D)$ over $\veck_D$ provides the total number
${\cal P}_D$ of QD pairs, and, consequently, allows for an {\sl ab initio}
calculation of the Levinger's factor, $L(A)$. The CBF results for
symmetrical nuclear matter, $L(\infty)=11.63$, is about $20\%$ larger
than $L_{Lev}(\infty)=9.26$ given in the literature. In the case of 
heavy nuclei $L_{Lev}(A)$ and  $L_{Laget}(A)$  bracket our CBF results,
which are therefore consistent with the available photoreaction data
within the quasideuteron model phenomenology.

It should be noticed that the theoretical estimate of $L(A)$ in the
range $150\leq A \leq 250$ is fairly constant, its increase
with $A$ being of $\sim 3\%$. The $A^{-1/3}$ surface behavior leads to a
very slow increase of $L(A)$ with $A$, and at $A\sim 200$ we are still
quite far away from the asymptotic region.

In addition,
the analysis described in this paper shows that when a deuteron is 
embedded in nuclear matter at equilibrium density, its wave function 
gets appreciably modified by the surrounding medium. While 
in the case of the S-wave component the difference is mostly visible at small 
relative distance ($r < 1$ fm), the D-wave component of the QD appears to be 
significantly quenched, with respect to the deuteron $w_D(r)$, over the range 
0 $< r <$ 2 fm. It has to be pointed out, however, that the radius of the 
QD configuration is very close to the deuteron radius, the difference being 
$\sim$ 2 $\%$. This result is in agreement with the conclusions of a 
recent study of deuteron-like configurations in light nuclei \cite{For96}. 
The authors of Ref.\cite{For96} find that the density distributions of $np$
pairs carrying the deuteron quantum numbers 
in $^3$He, $^4$He, $^6$Li, $^7$Li and
$^{16}$O exhibit size and structure similar to those observed in the deuteron.

The relative momentum distribution of a QD pair, $P^{\it rel}_D(k)$, 
extends into the region 
$|\veck| > k_F$, where it appears to be strongly suppressed 
with respect to the corresponding deuteron momentum distribution 
$|\Psi_D(k)|^2$, although $|\Psi_{QD}(k)|^2$ is larger than 
$|\Psi_D(k)|^2$ at
high $k$. It has to 
be pointed out that the behavior of $P^{\it rel}_D(k)$ at $k > k_F$ is entirely dictated 
by the high momentum tail of the nuclear matter 
momentum distribution, produced by 
strong short range NN correlations. Within the Fermi gas model $n(k>k_F) 
\equiv$ 0, and $P^{\it rel}_D(k>k_F)$ vanishes identically.

Higher order cluster terms, neglected in this paper and arising from the 
inclusion of additional bonds in the diagrammatical structure of 
fig. \ref{f1}, are not expected to 
change the main conclusions of the present paper, neither regarding 
the behavior of the deuteron distribution in nuclear matter, nor as far as 
the disussion on the Levinger's factor is concerned.

In view of the relevance that 
$P_D(\veck_D)$ and $|\Psi_{QD}(k)|^2$ may assume in the study of 
those lepton--nucleus reactions where the ejected hadron is in 
kinematical regions forbidden to lepton--nucleon processes, 
the calculations presented in this paper need to be extended 
 $i)$ by introducing higher order cluster terms in the expansion 
of the two--body density matrix,  and $ii)$ by 
explicitely considering finite nuclei wave functions. 
Work in these directions is in progress.

\acknowledgements

Two of us (O.B. and A.F.) thank the Institute for Nuclear Theory at the 
University of Washington for its hospitality and the Department of Energy
for partial support during the completion of this work. The support from the
International Centre for Theoretical Physics and 
from RFFI grants N98-02-17463 and N99-02-17727 is gratefully acknowledged 
by A.Yu.I. and G.I.L.


\begin{figure}
\caption{
Diagram showing the cluster contribution to $P_D(\veck_D)$ of 
eqs.(\protect\ref{def:P_D}) and (\protect\ref{def:P_D2}) 
considered in this paper.
The oriented solid lines represent the correlated one-body density matrix, 
whereas the wiggly lines correspond to the {\it dressed} 
deuteron-like $np$ pairs.
}
\label{f1}
\end{figure}

\begin{figure}
\caption{
Upper panel: the solid line shows the radial dependence of $U(r)$,  
defined by Eq.(\protect\ref{def:U-coord}), while the dashed and dot-dash
lines correspond to $u_D(r)$ and $\Delta u(r)$, respectively.
Lower panel: same as the upper panel, but for the $\ell$=2 components
of the QD and deuteron wave functions.
All wave functions are given in units of (GeV/c)$^{3/2}$.
}
\label{f2}
\end{figure}

\begin{figure}
\caption{
Same as in fig. \protect\ref{f2} in momentum space.
All wave functions are given in units of (GeV/c)$^{-3/2}$.
}
\label{f3}
\end{figure}

\begin{figure}
\caption{
Radial dependence of the functions $\Delta H_c(r)$ and $H_t(r)$, entering the 
definitions of $\Delta u(r)$ and $\Delta w(r)$ 
(see Eqs.(\ref{def:DU-coord}) and 
(\ref{def:DW-coord}).
}
\label{f4}
\end{figure}
\begin{figure}
\caption{
Momentum distribution of QD pairs in nuclear matter at 
equilibrium density as a function of the 
total momentum $|\veck_D|$ (see. Eqs.(\protect\ref{def:P_D})
and (\ref{def:P_D2})). Solid line: correlated model; 
dashed line: deuterons in a Fermi gas model.
The insert shows a blow up of the region $|\veck_D|/2k_F <$ 1, plotted in 
linear scale.
}
\label{f5}
\end{figure}

\begin{figure}
\caption{
The solid line shows the relative momentum distribution of a QD pair in 
nuclear matter at equilibrium density (Eq.(\protect\ref{def:Phid})).  
The dashed and dot-dash lines correspond to $\phi(k)$ 
(in units of (GeV/c)$^3$) and 
$|\Psi_{QD}(k)|^2$ (in units of (GeV/c)$^{-3}$), defined by 
Eqs.(\protect\ref{def:phi}) and (\protect\ref{PSID}). 
The diamonds show the squared 
momentum space wave function of a free deuteron.
}
\label{f6}
\end{figure}
\begin{figure}
\caption []{
The CBF Levinger's factor $L(A)$ of heavy nuclei (solid line) and 
nuclear matter (indicated by the arrow). 
The LDA approximation of ref. \cite{Fabrocini85} has been used 
for heavy nuclei. 
The phenomonological values of $L_{Lev}(A)$ corresponding to photoreaction 
data of Lepretre {\sl et al.} \cite{Lepretre81} (squares) and 
Ahrens {\sl et al.} \cite{Ahrens75} (crosses and diamonds) 
 are taken from ref.\cite{Ang86}. The empirical values of $L_{Lev}(A)$ 
represented by circles are from ref.\cite{Carlos81}. 
}
\label{f7}
\end{figure}

\begin{references}

\bibitem{Lev51}
J.S. Levinger, Phys. Rev. {\bf 84}, 43 (1951).

\bibitem{Got58}
K. Gottfried, Nucl. Phys. {\bf 5}, 557 (1958).

\bibitem{Lev79}
J.S. Levinger,  Phys. Lett. {\bf B82}, 181 (1979).

\bibitem{Laget81}
J.M. Laget, Nucl. Phys.  {\bf A358}, 275c (1981).

\bibitem{Frank81}
L.L. Frankfurt, and M.I. Strikman, Phys. Rep. {\bf 76}, 215 (1981);
{\bf 160}, 236 (1988).

\bibitem{For96}
J.L. Forest, V.R. Pandharipande, S.C. Pieper, R.B. Wiringa, R. Schiavilla, 
and A. Arriaga, Phys. Rev. C {\bf 54}, 646 (1996).

\bibitem{Marcelbook}
{\it Nuclear Methods and the Nuclear Equation of State}, edited by 
M. Baldo (World Scientific, Singapore, 1999).

\bibitem{Fan84}
S. Fantoni, and V.R. Pandharipande, Nucl. Phys. {\bf A427}, 473 (1984).

\bibitem{Ben89}
O. Benhar, S. Fantoni, and A. Fabrocini, Nucl. Phys. {\bf A505}, 267 (1989).

\bibitem{Ben92}
O. Benhar, S. Fantoni, and A. Fabrocini, Nucl. Phys. {\bf A550}, 201 (1992).

\bibitem{Ben94}
O. Benhar, A. Fabrocini, S. Fantoni, and I. Sick, Nucl. Phys. 
{\bf A579}, 493.
(1994).

\bibitem{Ben00}
O. Benhar, and A. Fabrocini, Phys. Rev. C {\bf 62}, 034304 (2000).

\bibitem{Pan79}
V.R. Pandharipande, and R.B. Wiringa, Rev. Mod. Phys. {\bf 51}, 821 (1979).

\bibitem{Ang86}
M. Anghinolfi, {\it et al.},  Nucl. Phys. {\bf A457}, 645 (1986).

\bibitem{Carlos82}
P. Carlos,  H.Beil, R. Berg\'{e}re, A. Lepretre, and
A. Veyssi\'{e}re, Nucl. Phys. {\bf A378}, 317 (1982).

\bibitem{Lepretre81}
A. Lepretre, H.Beil, R. Berg\'{e}re, P. Carlos, J. Fagot, A. de Miniac, and
A. Veyssi\'{e}re, Nucl. Phys. {\bf A367}, 237 (1981).

\bibitem{Ahrens75}
J.Ahrens, H. Barchert, K.H. Czock, H.B. Eppler, H. Gimm, H. Gundrum,
M. Kroning, P. Rihem, G. Sita Ram, A. Zieger, and B. Ziegler, Nucl. Phys.
{\bf A251}, 479 (1975).

\bibitem{Fan98}
S. Fantoni, and A. Fabrocini, in {\it Microscopic Quantum Many-Body Theories 
and their Applications}, edited by J. Navarro and A. Polls, Lecture Notes in 
Physics, Vol. 510 (Springer-Verlag, Berlin, 1998), p.119.

\bibitem{Lag81}
I.E. Lagaris, and V.R. Pandharipande, Nucl. Phys. {\bf A359}, 331 (1981).


\bibitem{Fabrocini85}
A. Fabrocini, I.E. Lagaris, M. Viviani, and S. Fantoni, Phys. Lett. {\bf 156B},
277 (1985).

\bibitem{Carlos81}
P. Carlos, Lecture Notes in Phys. {\bf 137}, 168 (1981).

\end{references}
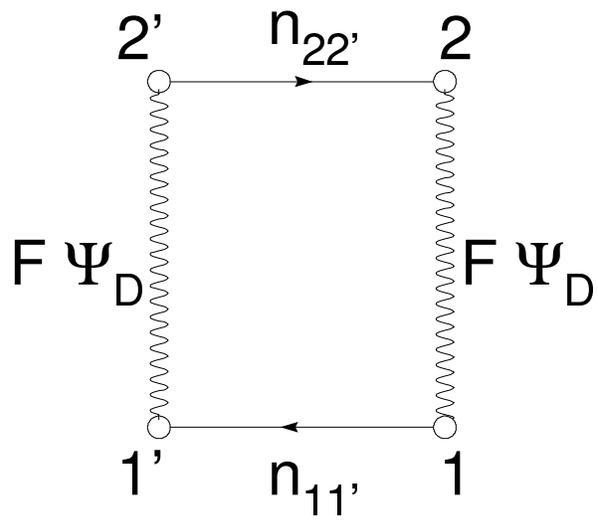
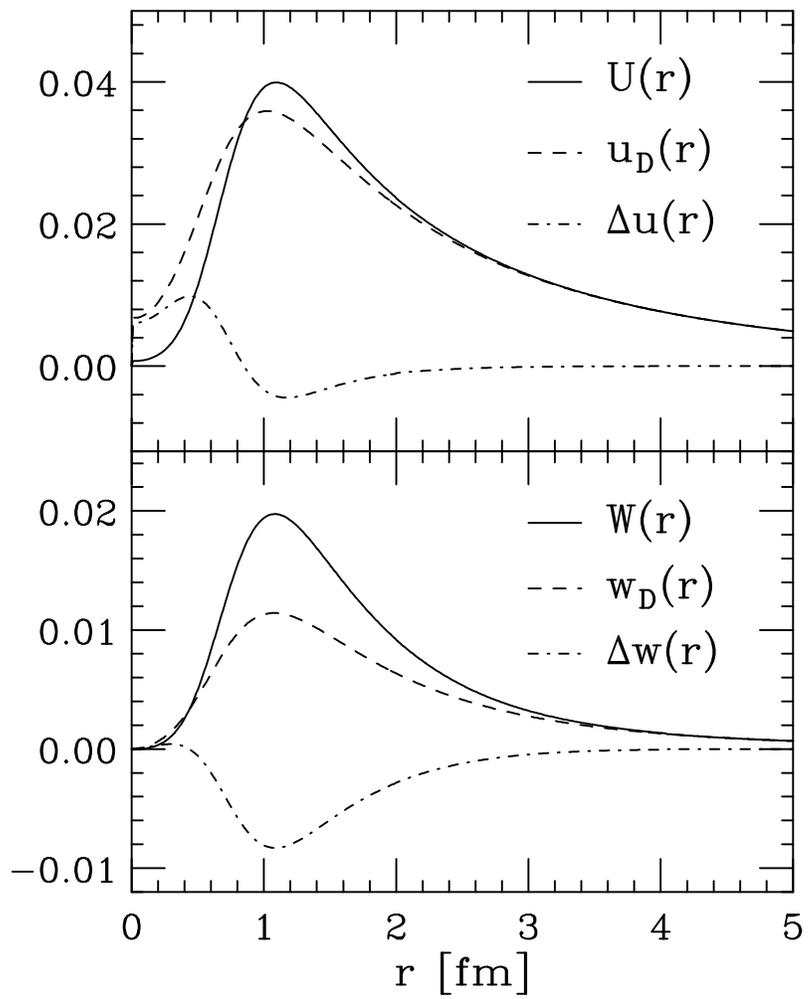
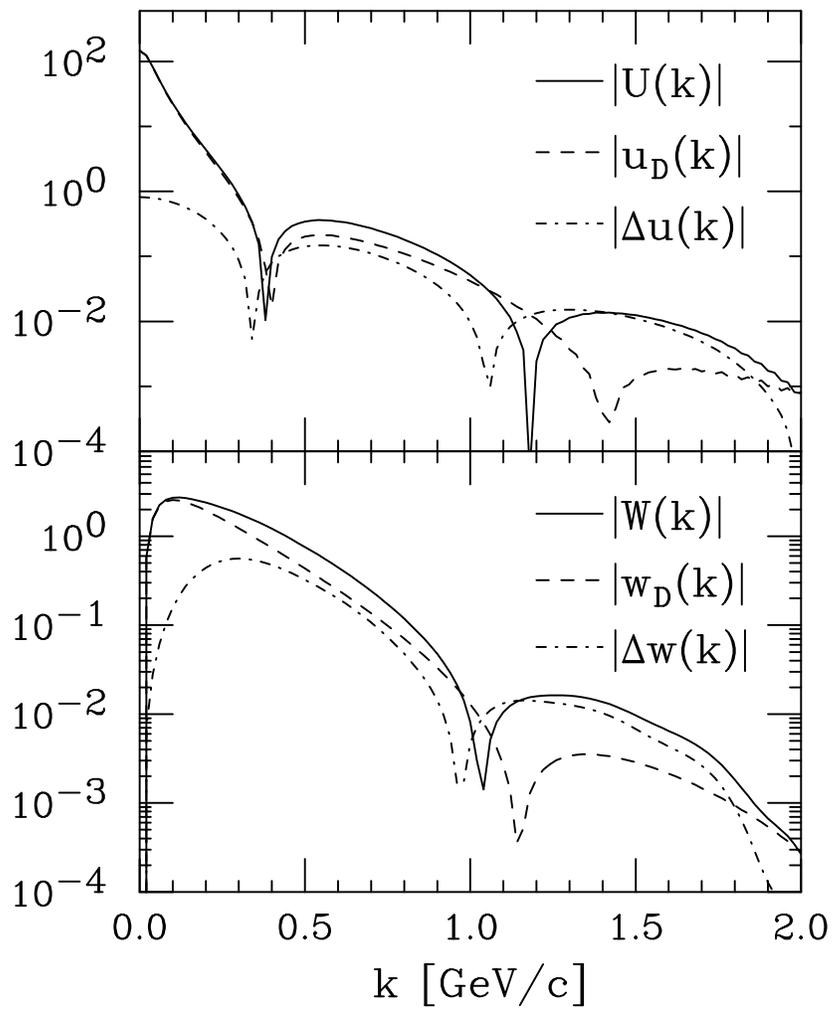
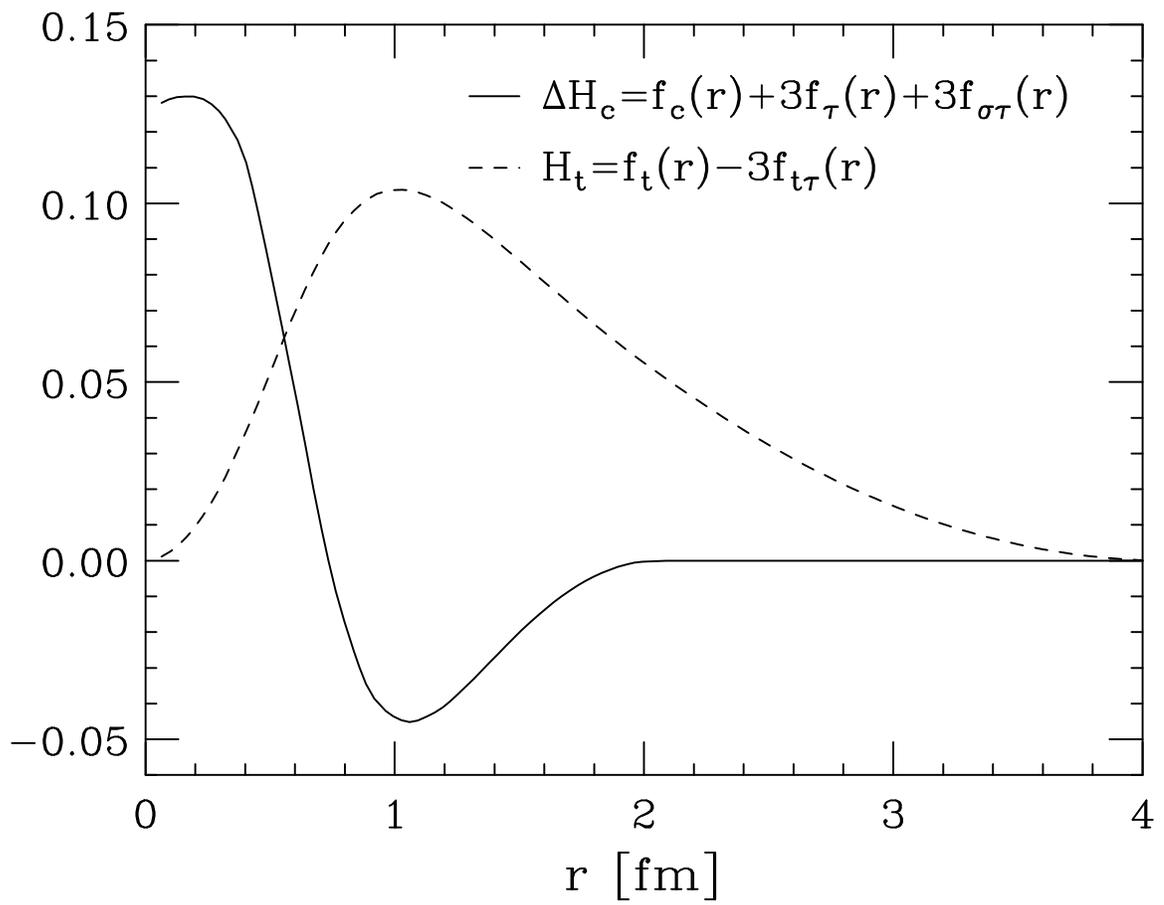
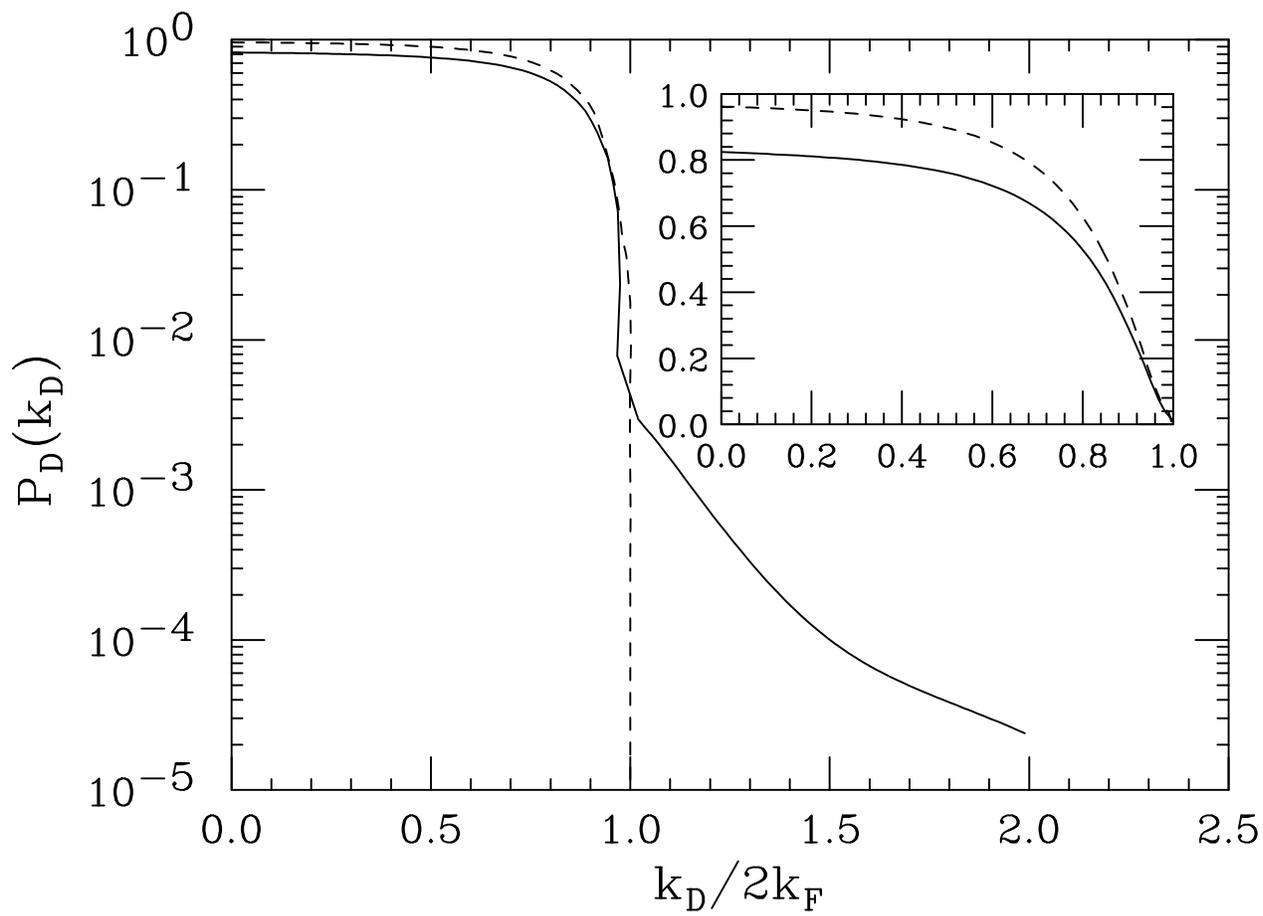
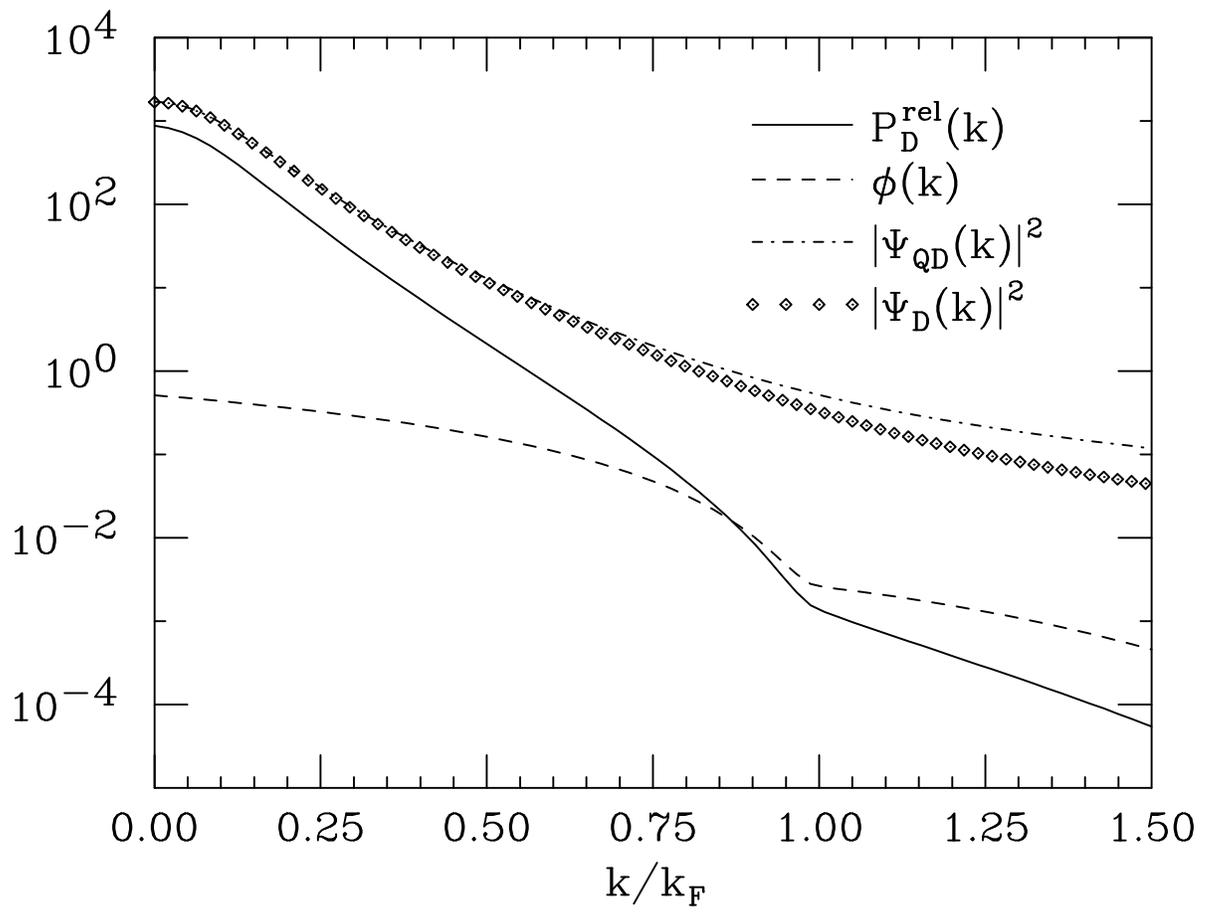
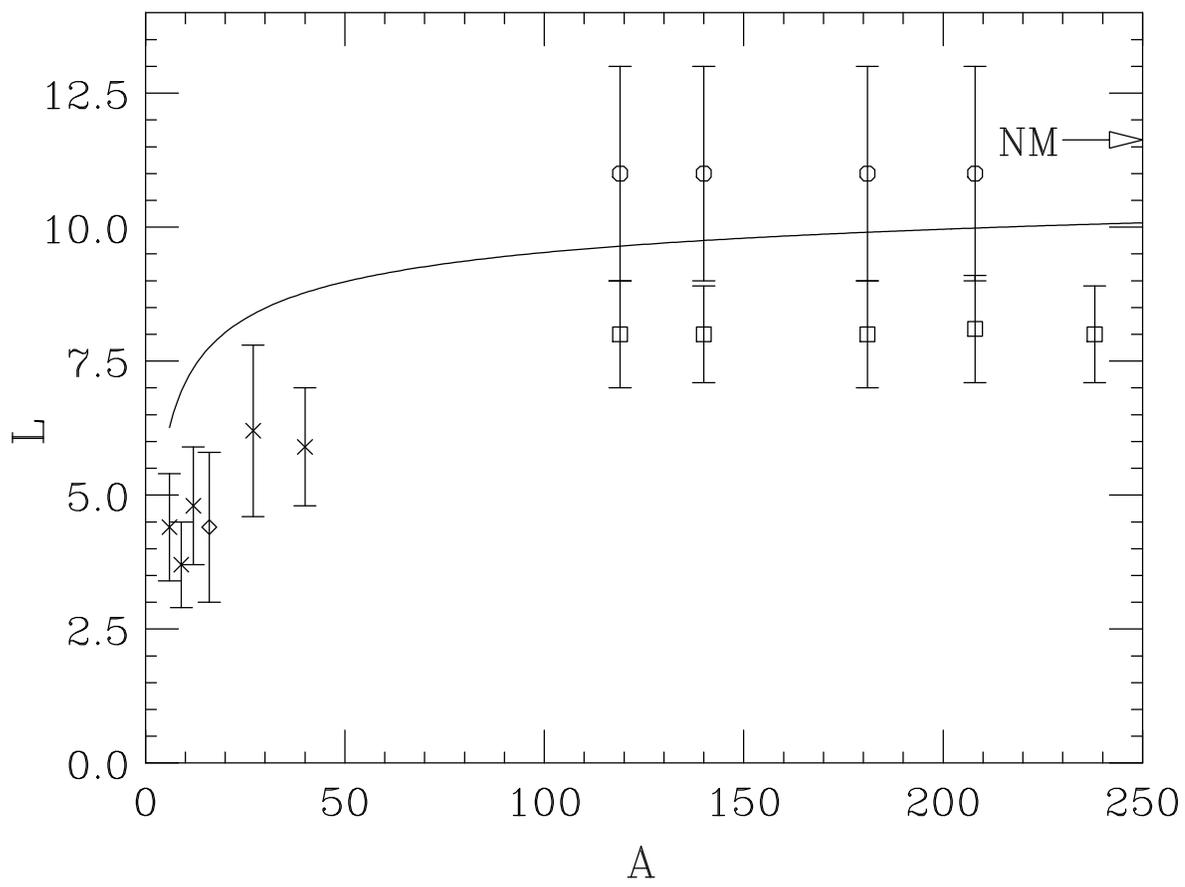
\end{document}